\documentclass[11pt]{article}
\usepackage{amssymb,amsmath,amsfonts}
\usepackage{graphicx}
\usepackage{graphics}
\usepackage{eepic,epsfig}

\begin{document}

\title{Scalar Casimir densities for cylindrically symmetric Robin boundaries}
\author{A. A. Saharian$^{1,2}$\thanks{%
E-mail: saharyan@server.physdep.r.am } \thinspace and A. S. Tarloyan$^{1}$ \\
\\
\textit{$^1$Department of Physics, Yerevan State University} \\
\textit{1 Alex Manoogian St, 375025 Yerevan, Armenia}\\
\textit{$^2$Departamento de F\'{\i}sica-CCEN, Universidade Federal da Para%
\'{\i}ba}\\
\textit{58.059-970, J. Pessoa, PB C. Postal 5.008, Brazil}}
\maketitle

\begin{abstract}
Wightman function, the vacuum expectation values of the field
square and the energy-momentum tensor are investigated for a
massive scalar field with general curvature coupling parameter in
the region between two coaxial cylindrical boundaries. It is
assumed that the field obeys general Robin boundary conditions on
bounding surfaces. The application of a variant of the generalized
Abel-Plana formula allows to extract from the expectation values
the contribution from single shells and to present the
interference part in terms of exponentially convergent integrals.
The vacuum forces acting on the boundaries are presented as the
sum of self--action and interaction terms. The first one contains
well-known surface divergences and needs a further
renormalization. The interaction forces between the cylindrical
boundaries are finite and are attractive for special cases of
Dirichlet and Neumann scalars. For the general Robin case the
interaction forces can be both attractive or repulsive depending
on the coefficients in the boundary conditions. The total Casimir
energy is evaluated by using the zeta function regularization
technique. It is shown that it contains a part which is located on
bounding surfaces. The formula for the interference part of the
surface energy is derived and the energy balance is discussed.
\end{abstract}

\bigskip


\bigskip

\section{Introduction}

\label{sec:introd}

The Casimir effect is one of the most interesting macroscopic manifestations
of the nontrivial structure of the vacuum state in quantum field theory. The
effect is a phenomenon common to all systems characterized by fluctuating
quantities and results from changes in the vacuum fluctuations of a quantum
field that occur because of the imposition of boundary conditions or the
choice of topology. It may have important implications on all scales, from
cosmological to subnuclear, and has become in recent decades an increasingly
popular topic in quantum field theory. In addition to its fundamental
interest the Casimir effect also plays an important role in the fabrication
and operation of nano- and micro-scale mechanical systems. The imposition of
boundary conditions on a quantum field leads to the modification of the
spectrum for the zero-point fluctuations and results in the shift in the
vacuum expectation values for physical quantities such as the energy density
and stresses. In particular, the confinement of quantum fluctuations causes
forces that act on constraining boundaries. The particular features of the
resulting vacuum forces depend on the nature of the quantum field, the type
of spacetime manifold, the boundary geometries and the specific boundary
conditions imposed on the field. Since the original work by Casimir \cite%
{Casi48} many theoretical and experimental works have been done on this
problem (see, e.g., \cite{Most97,Plun86,Bord01,Milt02} and references
therein). Many different approaches have been used: mode summation method
with combination of the zeta function regularization technique, Green
function formalism, multiple scattering expansions, heat-kernel series, etc.
Advanced field-theoretical methods have been developed for Casimir
calculations during the past years \cite{Bord96}.

An interesting property of the Casimir effect has always been the geometry
dependence. Straightforward computations of geometry dependencies are
conceptually complicated, since relevant information is subtly encoded in
the fluctuations spectrum. Analytic solutions can usually be found only for
highly symmetric geometries including planar, spherically and cylindrically
symmetric boundaries. Aside from their own theoretical and experimental
interest, the problems with this type of boundaries are useful for testing
the validity of various approximations used to deal with more complicated
geometries. In particular, a great deal of attention received the
investigations of quantum effects for cylindrical boundaries. In addition to
traditional problems of quantum electrodynamics under the presence of
material boundaries, the Casimir effect for cylindrical geometries can also
be important to the flux tube models of confinement \cite{Fish87,Barb90} and
for determining the structure of the vacuum state in interacting field
theories \cite{Ambj83}. The calculation of the vacuum energy of
electromagnetic field with boundary conditions defined on a cylinder turned
out to be technically a more involved problem than the analogous one for a
sphere. First the Casimir energy of an infinite perfectly conducting
cylindrical shell has been calculated in Ref. \cite{Dera81} by introducing
ultraviolet cutoff and later the corresponding result was derived by zeta
function technique \cite{Milt99,Gosd98,Lamb99}. The local characteristics of
the corresponding electromagnetic vacuum such as energy density and vacuum
stresses are considered in \cite{Sah1cyl} for the interior and exterior
regions of a conducting cylindrical shell, and in \cite{Sah2cyl} for the
region between two coaxial shells (see also \cite{Sahrev}). The vacuum
forces acting on the boundaries in the geometry of two cylinders are also
considered in Refs. \cite{Mazz02}. Less symmetric configurations of a
semi-circular infinite cylinder and of a wedge with a coaxial cylindrical
boundary are investigated in Refs. \cite{Nest01,Reza02}. A large number of
papers is devoted to the investigation of the various aspects of the Casimir
effect for a dielectric cylinder (see, for instance, \cite{CasDielec} and
references therein). From another perspective, the influence of a dielectric
cylinder on the radiation process from a charged particle has been discussed
in \cite{Saha05rad}.

An interesting topic in the investigations of the Casimir effect is the
dependence of the vacuum characteristics on the nature of boundary
conditions imposed. In Ref. \cite{Rome01} scalar vacuum densities and the
zero-point energy for general Robin condition on a cylindrical surface in
arbitrary number of spacetime dimensions are studied for massive scalar
field with general curvature coupling parameter. The Robin boundary
conditions are an extension of the ones imposed on perfectly conducting
boundaries and may, in some geometries, be useful for depicting the finite
penetration of the field into the boundary with the 'skin-depth' parameter
related to the Robin coefficient \cite{Most85,Lebe01}. It is interesting to
note that the quantum scalar field satisfying Robin condition on the
boundary of a cavity violates the Bekenstein's entropy-to-energy bound near
certain points in the space of the parameter defining the boundary condition
\cite{Solod}. This type of conditions also appear in considerations of the
vacuum effects for a confined charged scalar field in external fields \cite%
{Ambjorn2} and in quantum gravity \cite{Mo,EK,Esp97}. Mixed boundary
conditions naturally arise for scalar and fermion bulk fields in braneworld
models \cite{Gherg}. For scalar fields with Robin boundary conditions, in
Ref. \cite{Rome02} it has been shown that in the discussion of the relation
between the mode-sum energy, evaluated as the sum of the zero-point energies
for each normal mode of frequency, and the volume integral of the
renormalized energy density for the Robin parallel plates geometry it is
necessary to include in the energy a surface term concentrated on the
boundary (for the discussion of the relation between the local and global
characteristics of the vacuum see also Ref. \cite%
{Kenn80,Full03,Saha01,Rome01,Cave05}). An expression for the surface
energy-momentum tensor for a scalar field with a general curvature coupling
parameter in the general case of bulk and boundary geometries is derived in
Ref. \cite{Saha04}. Related cosmological constant induced on the brane by
quantum fluctuations of a bulk field in braneworld scenarios has been
considered in \cite{Saha04b}.

In the present paper, we consider the Casimir densities in the region
between two coaxial cylindrical shells on background of the $(D+1)$%
-dimensional Minkowski spacetime. The positive frequency Wightman function,
the vacuum expectation values of the field square and the energy-momentum
tensor are investigated for a massive scalar field with general curvature
coupling parameter. In addition to describing the physical structure of the
quantum field at a given point, the energy-momentum tensor acts as the
source of gravity in the Einstein equations. It therefore plays an important
role in modelling a self-consistent dynamics involving the gravitational
field \cite{Birr82}. For the general case of Robin boundary conditions with
different coefficients for the inner and outer boundaries, we derive
formulae for the forces acting on the boundaries due to the modification of
the spectrum of the zero-point fluctuations by the presence of the second
boundary. The Casimir energy and the surface energy are investigated as well
and the energy balance is discussed.

The plan of the paper is as follows. In the next section we derive a formula
for the Wightman function in the region between two cylindrical surfaces.
The reason for our choice of the Wightman function is that this function
also determines the response of the particle detectors in a given state of
motion. To evaluate the bilinear field products we use the mode-sum method
in combination with the summation formula from \cite{Sahrev,Saha01}. These
formula allows (i) to extract from vacuum expectation values the parts due
to a single cylindrical boundary, and (ii) to present the interference parts
in terms of exponentially convergent integrals involving the modified Bessel
functions. The vacuum expectation values of the field square and the
energy-momentum tensor are obtained from the Wightman function and are
investigated in section \ref{sec:vevemt}. The vacuum forces acting on the
bounding surfaces are considered in section \ref{sec:forces}. They are
presented as the sum of self--action and interaction terms. The formulae are
derived for the interaction forces between the cylinders. Section \ref%
{sec:toten} is devoted to the total vacuum energy evaluated as a sum of
zero-point energies of elementary oscillators. We show that this energy in
addition to the volume part contains a part located on the bounding
surfaces. The formula for the interaction part of the surface energy is
derived. Further we discuss the relation between the vacuum energies and
forces acting on the boundaries. Section \ref{sec:Conc} concludes the main
results of the paper.

\section{Wightman function}

\label{sec:WF}

Consider a real scalar field $\varphi $ with curvature coupling parameter $%
\xi $ satisfying the field equation
\begin{equation}
\left( \nabla ^{i}\nabla _{i}+\xi R+m^{2}\right) \varphi \left( x\right) =0,
\label{fieldeq}
\end{equation}%
where $R$ is the curvature scalar for a $(D+1)$-dimensional background
spacetime, $\nabla _{i}$ is the covariant derivative operator. For special
cases of minimally and conformally coupled scalars one has $\xi =0$ and $\xi
=\xi _{D}\equiv (D-1)/4D$, respectively. Our main interest in this paper
will be one-loop quantum vacuum effects induced by two infinitely long
coaxial cylindrical surfaces with radii $a$ and $b$, $a<b$, in the Minkowski
spacetime. For this problem the background spacetime is flat and in Eq. (\ref%
{fieldeq}) we have $R=0$. As a result the eigenmodes are independent on the
curvature coupling parameter. However, the local characteristics of the
vacuum such as energy density and vacuum stresses \ depend on this
parameter. In accordance with the problem symmetry we will use cylindrical
coordinates $\left( r,\phi ,z_{1},...,z_{N}\right) $, \ $N=D-2$, and will
assume that the field obeys Robin boundary conditions on bounding surfaces:%
\begin{equation}
\left( A_{j}+B_{j}n_{(j)}^{i}\nabla _{i}\right) \varphi \left( x\right) \big|%
_{r=j}=0,\;j=a,b,  \label{boundcond}
\end{equation}%
with $A_{j}$ \ and $B_{j}$ being constants, $n_{(j)}^{i}$ \ is the
inward-pointing normal to the bounding surface $r=j$. For the region between
the surfaces, $a\leqslant r\leqslant b$, one has $n_{(j)}^{i}=n_{j}\delta
_{1}^{i}$ with the notations $n_{a}=1$ and $n_{b}=-1$. The imposition of
boundary conditions on the quantum field modifies the spectrum for
zero-point fluctuations and leads to the modification of the vacuum
expectation values (VEVs) for physical quantities compared with the case
without boundaries. First we consider the positive frequency Wightman
function. The VEVs of the field square and the energy-momentum tensor can be
evaluated on the base of this function. By the same method described below
any other two-point function can be evaluated.

Let $\left\{ \varphi _{\alpha }(x),\varphi _{\alpha }^{\ast }(x)\right\} $\
\ is a complete orthonormal set of positive and negative frequency solutions
to the field equation, specified by a set of quantum numbers $\alpha $ and
satisfying the boundary conditions (\ref{boundcond}). By expanding the field
operator and using the standard commutation relations, the positive
frequency Wightman function is presented as the mode-sum%
\begin{equation}
\left\langle 0|\varphi (x)\varphi (x^{\prime })|0\right\rangle =\sum_{\alpha
}\varphi _{\alpha }(x)\varphi _{\alpha }^{\ast }(x),  \label{W1}
\end{equation}%
where $|0\rangle $ is the amplitude for the corresponding vacuum state. For
the region $a\leqslant r\leqslant b$ the eigenfunctions are specified by the
set of quantum numbers $\alpha =(n,\gamma ,\mathbf{k})$, $n=0,\pm 1,\pm
2,\ldots $, and have the form%
\begin{equation}
\varphi _{\alpha }(x)=\beta _{\alpha }g_{\left\vert n\right\vert }(\gamma
a,\gamma r)\exp (in\phi +i\mathbf{kr}_{\parallel }-i\omega t),
\label{eigfunc1}
\end{equation}%
with $\mathbf{r}_{\parallel }=\left( z_{1},...,z_{N}\right) $, $\omega =%
\sqrt{\gamma ^{2}+k_{m}^{2}}$, $k_{m}^{2}=k^{2}+m^{2}$, and
\begin{equation}
g_{n}(\gamma a,\gamma r)=\bar{Y}_{n}^{(a)}(\gamma a)J_{n}(\gamma r)-\bar{J}%
_{n}^{(a)}(\gamma a)Y_{n}(\gamma r).  \label{gn}
\end{equation}%
In Eq. (\ref{gn}), $J_{n}(z)$ and $Y_{n}(z)$ are the Bessel and Neumann
functions and for a given function $f(z)$ we use the notation%
\begin{equation}
\bar{f}^{(j)}(z)\equiv A_{j}f(z)+(n_{j}B_{j}/j)zf^{\prime }(z),  \label{fbar}
\end{equation}%
with $\ \ j=a,b$. The eigenfunctions (\ref{eigfunc1}) with the radial part (%
\ref{gn}) satisfy the boundary condition on the inner surface. The
eigenvalues for the quantum number $\gamma $ are quantized by the boundary
condition (\ref{boundcond}) on the surface $r=b$. From this condition it
follows that the possible values of $\gamma $ are solutions to the equation (%
$\eta =b/a$)%
\begin{equation}
C_{n}^{ab}(\eta ,\gamma a)\equiv \bar{J}_{n}^{(a)}(\gamma a)\bar{Y}%
_{n}^{(b)}(\gamma b)-\bar{Y}_{n}^{(a)}(\gamma a)\bar{J}_{n}^{(b)}(\gamma
b)=0.  \label{modeeq}
\end{equation}%
In the discussion below the corresponding positive roots we will denote by $%
\gamma a=\sigma _{n,l}$, $l=1,2,\ldots $, assuming that they are arranged in
the ascending order, $\sigma _{n,l}<\sigma _{n,l+1}$.

The coefficient $\beta _{\alpha }$ in (\ref{eigfunc1}) is determined from
the orthonormality condition for the eigenfunctions%
\begin{equation}
\int dV\varphi _{\alpha }(x)\varphi _{\alpha ^{\prime }}^{\ast }(x)=\frac{1}{%
2\omega }\delta _{nn^{\prime }}\delta _{ll^{\prime }}\delta (\mathbf{k-k}%
^{\prime }),  \label{normcond}
\end{equation}%
where the integration goes over the region between the cylindrical shells.
By making use of the standard integral for the cylindrical functions (see,
for instance, \cite{Prud86}), one finds%
\begin{equation}
\beta _{\alpha }^{2}=\frac{\pi ^{2}\gamma T_{n}^{ab}(\gamma a)}{4\omega
a(2\pi )^{D-1}},  \label{betalf}
\end{equation}%
with the notation%
\begin{equation}
T_{n}^{ab}(z)=z\left\{ \frac{\bar{J}_{n}^{(a)2}(z)}{\bar{J}_{n}^{(b)2}(\eta
z)}\left[ A_{b}^{2}+(\eta ^{2}z^{2}-n^{2})\frac{B_{b}^{2}}{b^{2}}%
-A_{a}^{2}-(z^{2}-n^{2})\frac{B_{a}^{2}}{a^{2}}\right] \right\} ^{-1}.
\label{Tnab}
\end{equation}

Substituting eigenfunctions (\ref{eigfunc1}) into the mode-sum formula (\ref%
{W1}), for the positive frequency Wightman function one finds%
\begin{eqnarray}
\left\langle 0|\varphi (x)\varphi (x^{\prime })|0\right\rangle  &=&\frac{\pi
^{2}}{2a}\int d^{N}\mathbf{k}\sideset{}{'}{\sum}_{n=0}^{\infty
}\sum_{l=1}^{\infty }\frac{zg_{n}(z,zr/a)g_{n}(z,zr^{\prime }/a)}{(2\pi
)^{D-1}\sqrt{z+k_{m}^{2}a^{2}}}T_{n}^{ab}(z)\big|_{z=\sigma _{n,l}}  \notag
\\
&&\times \cos \left[ n\left( \phi -\phi ^{\prime }\right) \right] \exp [-i%
\mathbf{k}\left( \mathbf{r}_{\parallel }-\mathbf{r}_{\parallel }^{\prime
}\right) -i\omega \left( t-t^{\prime }\right) ]  \label{W2}
\end{eqnarray}%
where the prime on the summation sign means that the summand with $n=0$
should be halved. As the expressions for the eigenmodes $\sigma _{n,l}$ are
not explicitly known, the form (\ref{W2}) for the Wightman function is
inconvenient. For the further evaluation of this VEV we apply to the sum
over $l$ the summation formula \cite{Sahrev,Saha01}
\begin{eqnarray}
\frac{\pi ^{2}}{2}\sum_{l=1}^{\infty }h(\sigma _{n,l})T_{n}^{ab}(\sigma
_{n,l}) &=&\int\limits_{0}^{\infty }\frac{h(x)dx}{\bar{J}_{n}^{(a)2}(x)+\bar{%
Y}_{n}^{(a)2}(x)}  \notag \\
&&-\frac{\pi }{4}\int\limits_{0}^{\infty }dx\,\Omega _{an}(x,\eta x)\left[
h(xe^{\pi i/2})+h(xe^{-\pi i/2})\right] ,  \label{Abel}
\end{eqnarray}%
where%
\begin{equation}
\Omega _{an}(x,y)=\frac{\bar{K}_{n}^{(b)}(y)/\bar{K}_{n}^{(a)}(x)}{\bar{K}%
_{n}^{(a)}(x)\bar{I}_{n}^{(b)}(y)-\bar{K}_{n}^{(b)}(y)\bar{I}_{n}^{(a)}(x)},
\label{Oma}
\end{equation}%
and $I_{n}(x)$, $K_{n}(x)$ are the Bessel modified functions. Here we have
assumed that all zeros for the function $C_{n}^{ab}(\eta ,z)$ are real. In
case of existence of purely imaginary zeros we have to include additional
residue terms on the left of formula (\ref{Abel}) (see \cite{Saha01}).
Formula (\ref{Abel}) is valid for functions $h(z)$ satisfying the condition $%
\left\vert h(z)\right\vert <\epsilon (x)e^{c\left\vert y\right\vert }$, $%
z=x+iy$, $c<2(\eta -1)$, for large values $|z|$, where $x^{2\delta
_{B_{a}0}-1}\epsilon (x)\rightarrow 0$ for $x\rightarrow \infty $, and the
condition $h(z)=o(z^{-1})$ for $z\rightarrow 0$.

For the evaluation of the Wightman function, as a function $h(x)$ we choose%
\begin{equation}
h(x)=\frac{xg_{n}(x,xr/a)g_{n}(x,xr^{\prime }/a)}{\sqrt{x^{2}+k_{m}^{2}a^{2}}%
}\exp [i\sqrt{x^{2}/a^{2}+k_{m}^{2}}(t^{\prime }-t)].  \label{hx}
\end{equation}%
The corresponding conditions are satisfied if $r+r^{\prime }+|t-t^{\prime
}|<2b$. In particular, this is the case in the coincidence limit $%
t=t^{\prime }$ for the region under consideration. Now we can see that the
application of formula (\ref{Abel}) allows to present the Wightman function
in the form%
\begin{eqnarray}
\left\langle 0|\varphi (x)\varphi (x^{\prime })|0\right\rangle  &=&\frac{1}{%
(2\pi )^{D-1}}\sideset{}{'}{\sum}_{n=0}^{\infty }\cos \left[ n\left( \phi
-\phi ^{\prime }\right) \right] \int d^{N}\mathbf{k}\,e^{i\mathbf{k}\left(
\mathbf{r}_{\parallel }-\mathbf{r}_{\parallel }^{\prime }\right) }  \notag \\
&&\times \Bigg\{\frac{1}{a}\int\limits_{0}^{\infty }dz\frac{h(z)}{\bar{J}%
_{n}^{(a)2}(z)+\bar{Y}_{n}^{(a)2}(z)}-\frac{2}{\pi }\int\limits_{k_{m}}^{%
\infty }dz\frac{x\Omega _{a\nu }(az,bz)}{\sqrt{z^{2}-k_{m}^{2}}}  \notag \\
&&\times G_{n}^{(a)}(az,zr)G_{n}^{(a)}(az,zr^{\prime })\cosh \left[
(t-t^{\prime })\sqrt{z^{2}-k_{m}^{2}}\right] \Bigg\},  \label{W3}
\end{eqnarray}%
with the notations (the notation with $j=b$ will be used below)%
\begin{equation}
G_{n}^{(j)}(x,y)=\bar{K}_{n}^{(j)}(x)I_{n}(y)-\bar{I}_{n}^{(j)}(x)K_{n}(y),%
\;j=a,b.  \label{Gnj}
\end{equation}

In the limit $b\rightarrow \infty $ the second term in figure braces on the
right of (\ref{W3}) vanishes, whereas the first term does not depend on $b$.
It follows from here that the part with the first term presents the Wightman
function in the region outside of a single cylindrical shell with radius $a$
(of course, this may also be seen by direct evaluation of the corresponding
Wightman function). To simplify this part we use the identity%
\begin{eqnarray}
\frac{g_{n}(z,zr/a)g_{n}(z,zr^{\prime }/a)}{\bar{J}_{n}^{(a)2}(z)+\bar{Y}%
_{n}^{(a)2}(z)} &=&J_{n}(zr/a)J_{n}(zr^{\prime }/a)-\frac{1}{2}%
\sum\limits_{\sigma =1}^{2}\frac{\bar{J}_{n}^{(a)}(z)}{\bar{H}_{n}^{(\sigma
a)}(z)}  \notag \\
&&\times H_{n}^{(\sigma )}(zr/a)H_{n}^{(\sigma )}(zr^{\prime }/a),
\label{iden1}
\end{eqnarray}%
with $H_{n}^{(\sigma )}(z)$, $\sigma =1,2$ being the Hankel functions.
Substituting this into the first integral in the figure braces of Eq. (\ref%
{W3}) we rotate the integration contour over $z$ by the angle $\pi /2$ for $%
\sigma =1$ and by the angle $-\pi /2$ for $\sigma =2$. Under the condition $%
r+r^{\prime }-|t-t^{\prime }|>2a$, the integrals over the arcs of the circle
with large radius vanish. The integrals over $(0,iak_{m})$ and $(0,-iak_{m})$
cancel out and after introducing the Bessel modified functions one obtains%
\begin{eqnarray}
\int\limits_{0}^{\infty }dz\frac{h(z)/a}{\bar{J}_{n}^{(a)2}(z)+\bar{Y}%
_{n}^{(a)2}(z)} &=&\int\limits_{0}^{\infty }dzz\frac{J_{n}(zr)J_{n}(zr^{%
\prime })}{\sqrt{z^{2}+k_{m}^{2}}}\exp [i\sqrt{z^{2}+k_{m}^{2}}(t^{\prime
}-t)]  \notag \\
&-&\frac{2}{\pi }\int\limits_{k_{m}}^{\infty }dz\,z\frac{\bar{I}%
_{n}^{(a)}(az)}{\bar{K}_{n}^{(a)}(az)}\frac{K_{n}(zr)K_{n}(zr^{\prime })}{%
\sqrt{z^{2}-k_{m}^{2}}}\cosh [\sqrt{z^{2}-k_{m}^{2}}(t^{\prime }-t)].
\label{int1}
\end{eqnarray}%
Substituting this into formula (\ref{W3}), the Wightman function is
presented in the form%
\begin{eqnarray}
\left\langle 0|\varphi (x)\varphi (x^{\prime })|0\right\rangle
&=&\left\langle \varphi (x)\varphi (x^{\prime })\right\rangle
^{(0)}+\left\langle \varphi (x)\varphi (x^{\prime })\right\rangle ^{(a)} \nonumber \\
&& - \frac{2^{2-D}}{\pi ^{D}}\sideset{}{'}{\sum}_{n=0}^{\infty
}\cos \left[
n\left( \phi -\phi ^{\prime }\right) \right] \int d^{N}\mathbf{k}\,e^{i%
\mathbf{k}\left( \mathbf{r}_{\parallel }-\mathbf{r}_{\parallel }^{\prime
}\right) } \int\limits_{k_{m}}^{\infty }dz\,z\frac{\Omega _{an}(az,bz)}{\sqrt{%
z^{2}-k_{m}^{2}}} \notag \\
&& \times G_{n}^{(a)}(az,zr)G_{n}^{(a)}(az,zr^{\prime })\cosh
\left[ (t-t^{\prime })\sqrt{z^{2}-k_{m}^{2}}\right] ,  \label{W4}
\end{eqnarray}%
where $\left\langle \varphi (x)\varphi (x^{\prime })\right\rangle ^{(0)}$ is
the Wightman function for a scalar field in\ the unbounded Minkowskian
spacetime, and
\begin{eqnarray}
\left\langle \varphi (x)\varphi (x^{\prime })\right\rangle ^{(a)} &=&-\frac{%
2^{2-D}}{\pi ^{D}}\sideset{}{'}{\sum}_{n=0}^{\infty }\cos \left[ n\left(
\phi -\phi ^{\prime }\right) \right] \int d^{N}\mathbf{k}\,e^{i\mathbf{k}%
\left( \mathbf{r}_{\parallel }-\mathbf{r}_{\parallel }^{\prime }\right) }
\notag \\
&&\times \int\limits_{k_{m}}^{\infty }dz\,z\frac{\bar{I}_{n}^{(a)}(az)}{\bar{%
K}_{n}^{(a)}(az)}\frac{K_{n}(zr)K_{n}(zr^{\prime })}{\sqrt{z^{2}-k_{m}^{2}}}%
\cosh [\sqrt{z^{2}-k_{m}^{2}}(t^{\prime }-t)]  \label{Wa}
\end{eqnarray}%
is the part of the Wightman function induced by a single cylindrical shell
with radius $a$ in the region $r>a$. Hence, the last term on the right of (%
\ref{W4}) is induced by the presence of the second shell with radius $b$.

By using the identity%
\begin{eqnarray}
&&\sum_{j=a,b}n_{j}\Omega
_{jn}(az,bz)G_{n}^{(j)}(az,zr)G_{n}^{(j)}(az,zr^{\prime })  \notag \\
&=&\frac{\bar{K}_{n}^{(b)}(bz)}{\bar{I}_{n}^{(b)}(bz)}I_{n}(zr)I_{n}(zr^{%
\prime })-\frac{\bar{I}_{n}^{(a)}(az)}{\bar{K}_{n}^{(a)}(az)}%
K_{n}(zr)K_{n}(zr^{\prime }),  \label{iden2}
\end{eqnarray}%
with the notation%
\begin{equation}
\Omega _{bn}(x,y)=\frac{\bar{I}_{n}^{(a)}(x)/\bar{I}_{n}^{(b)}(y)}{\bar{K}%
_{n}^{(a)}(x)\bar{I}_{n}^{(b)}(y)-\bar{K}_{n}^{(b)}(y)\bar{I}_{n}^{(a)}(x)},
\label{Omb}
\end{equation}%
the Wightman function can also be presented in the equivalent form%
\begin{eqnarray}
\left\langle 0|\varphi (x)\varphi (x^{\prime })|0\right\rangle
&=&\left\langle \varphi (x)\varphi (x^{\prime })\right\rangle
^{(0)}+\left\langle \varphi (x)\varphi (x^{\prime })\right\rangle
^{(b)} \nonumber \\
&& -\frac{2^{2-D}}{\pi ^{D}}\sideset{}{'}{\sum}_{n=0}^{\infty
}\cos \left[
n\left( \phi -\phi ^{\prime }\right) \right] \int d^{N}\mathbf{k}\,e^{i%
\mathbf{k}\left( \mathbf{r}_{\parallel }-\mathbf{r}_{\parallel }^{\prime
}\right) } \int\limits_{k_{m}}^{\infty }dz\,z\frac{\Omega _{bn}(az,bz)}{\sqrt{%
z^{2}-k_{m}^{2}}} \notag \\
&& \times G_{n}^{(b)}(bz,zr)G_{n}^{(b)}(bz,zr^{\prime })\cosh
\left[ (t-t^{\prime })\sqrt{z^{2}-k_{m}^{2}}\right] ,  \label{W5}
\end{eqnarray}%
where%
\begin{eqnarray}
\left\langle \varphi (x)\varphi (x^{\prime })\right\rangle ^{(b)} &=&-\frac{%
2^{2-D}}{\pi ^{D}}\sideset{}{'}{\sum}_{n=0}^{\infty }\cos \left[ n\left(
\phi -\phi ^{\prime }\right) \right] \int d^{N}\mathbf{k}\,e^{i\mathbf{k}%
\left( \mathbf{r}_{\parallel }-\mathbf{r}_{\parallel }^{\prime }\right) }
\notag \\
&&\times \int\limits_{k_{m}}^{\infty }dz\,z\frac{\bar{K}_{n}^{(b)}(bz)}{\bar{%
I}_{n}^{(b)}(bz)}\frac{I_{n}(zr)I_{n}(zr^{\prime })}{\sqrt{z^{2}-k_{m}^{2}}}%
\cosh [\sqrt{z^{2}-k_{m}^{2}}(t^{\prime }-t)]  \label{Wb}
\end{eqnarray}%
is the part induced by a single cylindrical shell with radius $b$ in the
region $r<b$. Note that formulae (\ref{Wa}) and (\ref{Wb}) are related by
the interchange $a\rightleftarrows b$, $I_{n}\rightleftarrows K_{n}$. In the
formulae for the Wightman function given above the integration over the
angular part of the vector $\mathbf{k}$ can be done with the help of the
formula%
\begin{equation}
\int d^{N}{\mathbf{k}}\,\frac{e^{i{\mathbf{kx}}}F(k)}{(2\pi )^{\frac{N}{2}}}%
=\int_{0}^{\infty }dk\,k^{N-1}F(k)\frac{J_{N/2-1}(k|{\mathbf{x}}|)}{(k|{%
\mathbf{x}}|)^{N/2-1}},  \label{intformwf}
\end{equation}%
for a given function $F(k)$.

\section{ VEVs of the field square and the energy-momentum tensor}

\label{sec:vevemt}

Having the Wightman function we can evaluate the VEVs of the field square
and the energy-momentum tensor. These VEVs in the regions $r<a$ and $r>b$
are the same as those for a single cylindrical surface with radius $a$ and $%
\ b$ respectively and are investigated in \cite{Rome01}. For this reason in
the discussion below we will be concerned on the region $a<r<b$. By making
use of formulae (\ref{W4}) and (\ref{W5}) for the Wightman function and
taking the coincidence limit of the arguments, for the VEV of the field
square one finds%
\begin{equation}
\langle 0|\varphi ^{2}|0\rangle =\langle \varphi ^{2}\rangle ^{(0)}+\langle
\varphi ^{2}\rangle ^{(j)}-A_{D}\sideset{}{'}{\sum}_{n=0}^{\infty
}\int_{m}^{\infty }du\,u\left( u^{2}-m^{2}\right) ^{\frac{D-3}{2}}\Omega
_{jn}(au,bu)G_{n}^{(j)2}(ju,ru),  \label{phi21}
\end{equation}%
where $j=a$ and $j=b$ provide two equivalent representations and
\begin{equation}
A_{D}=\frac{2^{2-D}}{\pi ^{\frac{D+1}{2}}\Gamma \left( \frac{D-1}{2}\right) }%
.  \label{AD}
\end{equation}%
To obtain this result we have employed the integration formula%
\begin{equation}
\int_{0}^{\infty }dk\,\,\int\limits_{k_{m}}^{\infty }dz\,\frac{k^{D-3}f(z)}{%
\sqrt{z^{2}-k_{m}^{2}}}=\frac{\sqrt{\pi }\Gamma \left( \frac{D}{2}-1\right)
}{2\Gamma \left( \frac{D-1}{2}\right) }\int_{m}^{\infty }du\,\left(
u^{2}-m^{2}\right) ^{\frac{D-3}{2}}f(u).  \label{intform1}
\end{equation}%
For points away from the boundaries the last two terms on the right of
formula (\ref{phi21}) are finite and, hence, the subtraction of the
Minkowskian part without boundaries is sufficient to obtain the renormalized
value for the VEV: $\langle \varphi ^{2}\rangle _{\mathrm{ren}}=\langle
0|\varphi ^{2}|0\rangle -\langle \varphi ^{2}\rangle ^{(0)}$. In formula (%
\ref{phi21}) the part $\langle \varphi ^{2}\rangle ^{(j)}$ is induced by a
single cylindrical surface with radius $j$ when the second surface is
absent. The formulae for these terms are obtained from (\ref{Wa}) and (\ref%
{Wb}) in the coincidence limit. For $j=a$ one has
\begin{equation}
\langle \varphi ^{2}\rangle ^{(a)}=-A_{D}\sideset{}{'}{\sum}_{n=0}^{\infty
}\int_{m}^{\infty }du\,u\left( u^{2}-m^{2}\right) ^{\frac{D-3}{2}}\frac{\bar{%
I}_{n}^{(a)}(au)}{\bar{K}_{n}^{(a)}(au)}K_{n}^{2}(ru),  \label{phi2a}
\end{equation}%
and the formula for $\langle \varphi ^{2}\rangle ^{(b)}$ is obtained from
here by the replacements $a\rightarrow b$, $I\rightleftarrows K$. The last
term on the right of formula (\ref{phi21}) is induced by the presence of the
second cylindrical surface. The surface divergences in the boundary induced
parts in the VEV\ of the field square are the same as those for single
surfaces and are investigated in \cite{Rome01}. In particular, the last term
on the right of (\ref{phi21}) is finite for $r=j$. It follows from here that
if we present the renormalized VEV of the field square in the form%
\begin{equation}
\langle \varphi ^{2}\rangle _{\mathrm{ren}}=\sum_{j=a,b}\langle \varphi
^{2}\rangle ^{(j)}+\langle \varphi ^{2}\rangle ^{(ab)},  \label{interphi2}
\end{equation}%
then the interference part $\langle \varphi ^{2}\rangle ^{(ab)}$ is finite
on both boundaries. In the limit $a\rightarrow 0$ this part vanishes as $a$
for Robin boundary condition on the inner shell ($A_{a},B_{a}\neq 0$), as $%
a^{2}$ for Neumann boundary condition ($A_{a}=0$), and like $1/\ln a$ for
Dirichlet boundary condition ($B_{a}=0$). In the limit $b\rightarrow \infty $
and for a massless field the interference part vanishes as $\ln b/b^{D-1}$
for Robin boundary condition on the inner shell, as $\ln b/b^{D+1}$ for
Neumann boundary condition, and like $1/b^{D-1}$ for Dirichlet boundary
condition. In the same limit under the condition $mb\gg 1$ the interference
part is exponentially suppressed.

The VEV for the energy-momentum tensor is obtained by using the formulae for
the Wightman function and the VEV of the field square:%
\begin{equation}
\langle 0|T_{ik}|0\rangle =\lim_{x^{\prime }\rightarrow x}\partial
_{i}\partial _{k}^{\prime }\langle 0|\varphi (x)\varphi (x^{\prime
})|0\rangle +\left[ \left( \xi -\frac{1}{4}\right) g_{ik}\nabla _{l}\nabla
^{l}-\xi \nabla _{i}\nabla _{k}\right] \langle 0|\varphi ^{2}|0\rangle .
\label{emtvev1}
\end{equation}%
Substituting (\ref{W4}), (\ref{W5}), (\ref{phi21}) into (\ref{emtvev1}) one
finds (no summation over $i$)
\begin{eqnarray}
\langle 0|T_{i}^{k}|0\rangle &=&\langle T_{i}^{k}\rangle ^{(0)}+\langle
T_{i}^{k}\rangle ^{(j)}+A_{D}\delta _{i}^{k}\sideset{}{'}{\sum}%
_{n=0}^{\infty }\int_{m}^{\infty }du\,u^{3}  \notag \\
&&\times \left( u^{2}-m^{2}\right) ^{\frac{D-3}{2}}\Omega
_{jn}(au,bu)F_{n}^{(i)}[G_{n}^{(j)}(ju,ru)],  \label{vevemt2}
\end{eqnarray}%
where we have introduced notations
\begin{subequations}
\begin{eqnarray}
F_{n}^{(0)}[f(z)] &=&\frac{1-m^{2}r^{2}/z^{2}}{D-1}f^{2}(z)+\left( 2\xi -%
\frac{1}{2}\right) \left[ f^{\prime 2}(z)+\left( \frac{n^{2}}{z^{2}}%
+1\right) f^{2}(z)\right] ,  \label{Fn0} \\
F_{n}^{(1)}[f(z)] &=&\frac{1}{2}\left[ f^{\prime 2}(z)-\left( \frac{n^{2}}{%
z^{2}}+1\right) f^{2}(z)\right] +\frac{2\xi }{z}f(z)f^{\prime }(z),
\label{Fn1} \\
F_{n}^{(2)}[f(z)] &=&\left( 2\xi -\frac{1}{2}\right) \left[ f^{\prime
2}(z)+\left( \frac{n^{2}}{z^{2}}+1\right) f^{2}(z)\right] -\frac{2\xi }{z}%
f(z)f^{\prime }(z)+\frac{n^{2}}{z^{2}}f^{2}(z),  \label{Fn3}
\end{eqnarray}
with $f(z)=G_{n}^{(j)}(ju,z)$ and
\end{subequations}
\begin{equation}
F_{n}^{(i)}[f(z)]=F_{n}^{(0)}[f(z)],\quad i=3,\ldots ,D-1.  \label{Fni}
\end{equation}%
In formula (\ref{vevemt2}) the term $\langle T_{i}^{k}\rangle ^{(j)}$ is
induced by a single cylindrical surface with radius $j$. These parts for
both interior and exterior regions are investigated in \cite{Rome01}. The
formula for the case $j=a$ is obtained from (\ref{phi2a}) by the replacement
$K_{n}^{2}(ru)\rightarrow u^{2}F_{n}^{(i)}[K_{n}(ru)]$. As in the case of
the field square, the renormalized VEV\ of the energy-momentum tensor can be
presented in the form%
\begin{equation}
\langle T_{i}^{k}\rangle _{\mathrm{ren}}=\sum_{j=a,b}\langle
T_{i}^{k}\rangle ^{(j)}+\langle T_{i}^{k}\rangle ^{(ab)},  \label{Tikren}
\end{equation}%
where the surface divergences are contained in the single boundary parts
only and interference part is finite on the boundaries. The explicit formula
for the latter is obtained by subtracting from the last term on the right (%
\ref{vevemt2}) the corresponding single surface part. Two equivalent
representations are obtained by taking in (\ref{vevemt2}) $j=a$ or $\ j=b$.
Due to the presence of boundaries the vacuum stresses in radial, azimuthal
and axial directions are anisotropic. For the axial stress and the energy
density we have standard relation for the unbounded vacuum. It can be easily
checked that the separate terms in formulae (\ref{phi21}) and (\ref{vevemt2}%
) satisfy the standard trace relation%
\begin{equation}
T_{i}^{i}=D(\xi -\xi _{D})\nabla _{i}\nabla ^{i}\varphi ^{2}+m^{2}\varphi
^{2},  \label{trrel}
\end{equation}%
and the continuity equation $\nabla _{i}T_{k}^{i}=0$. For the geometry under
consideration the latter takes the form%
\begin{equation}
\frac{d}{dr}T_{1}^{1}+\frac{1}{r}\left( T_{1}^{1}-T_{2}^{2}\right) =0.
\label{conteq}
\end{equation}%
In particular, this means that the $r$-dependence of the radial pressure
leads to the anisotropy of the vacuum stresses. For a conformally coupled
massless scalar the vacuum energy-momentum tensor is traceless. In the limit
$a,b\rightarrow \infty $ with fixed $b-a$ from the formulae above the
results for the geometry of two parallel plates with Robin boundary
conditions are obtained. In the limit $a\rightarrow 0$ and for $A_{a}\neq 0$
the main contribution into the interference part of the vacuum
energy-momentum tensor comes from $n=0$ term and this part behaves like $a$
in the case $B_{a}\neq 0 $ and like $1/\ln (a/b)$ in the case $B_{a}=0$. In
the same limit and for $A_{a}=0$ the interference part behaves as $a^{2}$.
For a massless scalar field in the limit $b\rightarrow \infty $ and for $%
A_{a},B_{a}\neq 0$ the main contribution comes, again, from $n=0$ term. The
corresponding energy density behaves as $b^{1-D}$, while the vacuum stresses
behave like $\ln b/b^{D-1}$. For a massive scalar field under the condition $%
mb\gg 1$ the interference part of the energy-momentum tensor is
exponentially suppressed.

The interference part of the vacuum energy-momentum tensor gives finite
contribution into the vacuum energy in the region $a\leqslant r\leqslant b$.
To evaluate this contribution we note that the function $G_{n}^{(j)}(ju,ru)$
is a modified cylindrical function with respect to the argument $r$. By
using the formula for the integral involving the square of a modified
cylindrical function (see, for instance, \cite{Prud86}) and noting that for
these functions
\begin{equation}
z\left[ f^{\prime 2}(z)+\left( \frac{n^{2}}{z^{2}}+1\right) f^{2}(z)\right] =%
\left[ zf(z)f^{\prime }(z)\right] ^{\prime },  \label{relcylfunc}
\end{equation}%
the following formula can be obtained%
\begin{equation}
2\pi \int_{a}^{b}dr\,r\langle T_{0}^{0}\rangle ^{(ab)}=E_{b\leqslant
r<\infty }^{(a)}+E_{0\leqslant r\leqslant a}^{(b)}+\Delta E^{\mathrm{(vol)}},
\label{intvolen}
\end{equation}%
where $E_{b\leqslant r<\infty }^{(a)}$ ($E_{0\leqslant r\leqslant a}^{(b)}$)
is the vacuum energy in the region $b\leqslant r<\infty $ \ ($0\leqslant
r\leqslant a$) for a single cylindrical shell with radius $a$ ($b$) and%
\begin{eqnarray}
\Delta E^{\mathrm{(vol)}} &=&\pi A_{D}\sideset{}{'}{\sum}_{n=0}^{\infty
}\int_{m}^{\infty }du\,u\left( u^{2}-m^{2}\right) ^{\frac{D-3}{2}%
}\sum_{j=a,b}n_{j}\Omega _{jn}(au,bu)  \notag \\
&&\times \left\{ (4\xi -1)n_{j}A_{j}\frac{B_{j}}{j}-\frac{1-m^{2}/u^{2}}{D-1}%
\left[ B_{j}^{2}(u^{2}+n^{2}/j^{2})-A_{j}^{2}\right] \right\} .
\label{DeltaEvol}
\end{eqnarray}%
Now, by taking into account formula (\ref{intvolen}), for the total volume
energy in the region $a\leqslant r\leqslant b$ one finds%
\begin{equation}
E_{a\leqslant r\leqslant b}^{\mathrm{(vol)}}=2\pi \int_{a}^{b}dr\,r\langle
T_{0}^{0}\rangle _{\mathrm{ren}}=E_{r\geqslant a}^{(a,\mathrm{vol}%
)}+E_{r\leqslant b}^{(b,\mathrm{vol})}+\Delta E^{\mathrm{(vol)}},
\label{Evoln}
\end{equation}%
where $E_{r\geqslant a}^{(a,\mathrm{vol})}$ ($E_{r\leqslant b}^{(b,\mathrm{%
vol})}$) is the volume part of the vacuum energy outside (inside) a single
cylindrical shell with radius $a$ ($b$). Of course, due to the surface
divergences these single shell parts cannot be obtained directly by the
integration of the corresponding densities and need additional
renormalization. As we will see below in section \ref{sec:toten}, the total
vacuum energy in addition to the volume part contains the contribution
located on the boundaries.

\section{Interaction forces}

\label{sec:forces}

Now we turn to the vacuum forces acting on the cylindrical surfaces. The
vacuum force per unit surface of the cylinder at $r=j$ is determined by the $%
{}_{1}^{1}$ -- component of the vacuum energy-momentum tensor at this point.
For the region between two surfaces the corresponding effective pressures
can be presented as the sum of two terms:%
\begin{equation}
p^{(j)}=p_{1}^{(j)}+p_{\mathrm{(int)}}^{(j)},\quad j=a,b.  \label{pj}
\end{equation}%
The first term on the right is the pressure for a single cylindrical surface
at $r=j$ when the second surface is absent. This term is divergent due to
the surface divergences in the vacuum expectation values and needs further
renormalization. This can be done, for example, by applying the generalized
zeta function technique to the corresponding mode-sum. This procedure is
similar to that used in Ref. \cite{Rome01} for the evaluation of the total
Casimir energy for a single cylindrical shell. This calculation lies on the
same line with the evaluation of the surface Casimir densities and will be
presented in the forthcoming paper \cite{SahTarl}. Here we note that the
corresponding quantities will depend on the renormalization scale and can be
fixed by imposing suitable renormalization conditions which relates it to
observables. The second term on the right of Eq. (\ref{pj}) is the pressure
induced by the presence of the second cylinder, and can be termed as an
interaction force. Unlike to the single shell parts, this term is free from
renormalization ambiguities and is determined by the last term on the right
of formula (\ref{vevemt2}). Substituting into this term $r=j$ and using the
relations
\begin{equation}
G_{n}^{(j)}(u,u)=-n_{j}B_{j}/j,\quad G_{n}^{(j)^{\prime }}(u,u)=A_{j}/u,
\label{Gnuu}
\end{equation}%
for the interaction parts of the vacuum forces per unit surface one finds%
\begin{eqnarray}
p_{\mathrm{(int)}}^{(j)} &=&\frac{A_{D}}{2j^{2}}\sideset{}{'}{\sum}%
_{n=0}^{\infty }\int_{m}^{\infty }du\,u\left( u^{2}-m^{2}\right) ^{\frac{D-3%
}{2}}\Omega _{jn}(au,bu)  \notag \\
&&\times \left[ \left( n^{2}/j^{2}+u^{2}\right) B_{j}^{2}+4\xi
n_{j}A_{j}B_{j}/j-A_{j}^{2}\right] .  \label{pjint0}
\end{eqnarray}%
The expression on the right of this formula is finite for all non-zero
distances between the shells. By taking into account the inequalities $%
K_{n}(u)I_{n}(v)-K_{n}(v)I_{n}(u)>0$ and $K_{n}^{\prime }(u)I_{n}^{\prime
}(v)-K_{n}^{\prime }(v)I_{n}^{\prime }(u)<0$ for $u<v$, it can be seen that
the vacuum effective pressures are negative for both Dirichlet and Neumann
scalars and, hence, the corresponding interaction forces are attractive. For
the general Robin case the interaction force can be either attractive or
repulsive in dependence on the coefficients in the boundary conditions (see
the numerical example presented below in figure \ref{fig2}). The quantity $%
p_{\mathrm{(int)}}^{(j)}$ determines the force by which the scalar vacuum
acts on the cylindrical shell due to the modification of the spectrum for
the zero-point fluctuations by the presence of the second cylinder. As the
vacuum properties depend on the radial coordinate, there is no a priori
reason for the interaction terms (and also for the total pressures $p^{(j)}$%
) to be the same for $j=a$ and $\ j=b$, and the corresponding forces in
general are different. Note that the interaction parts act on the surfaces $%
r=a+0$ and $r=b-0$. The vacuum forces acting on the sides $r=a-0$ and $r=b+0$
are the same as those for single surfaces. In combination with the parts $%
p_{1}^{(j)}$ from (\ref{pj}), the latter give the total vacuum forces acting
on a single cylindrical shell. For Dirichlet and Neumann boundary conditions
these forces can be obtained by differentiation of the corresponding Casimir
energy (see below section \ref{sec:toten}). For $D=3$ massless scalar these
forces are repulsive for Dirichlet case and are attractive for Neumann case
\cite{Gosd98}.

Using the Wronskian for the Bessel modified functions, it can be seen that
for $j=a,b$ one has%
\begin{equation}
B_{jn}(u)\Omega _{jn}(au,bu)=jn_{j}\frac{\partial }{\partial j}\ln
\left\vert 1-\frac{\bar{I}_{n}^{(a)}(au)\bar{K}_{n}^{(b)}(bu)}{\bar{I}%
_{n}^{(b)}(bu)\bar{K}_{n}^{(a)}(au)}\right\vert ,  \label{ident2}
\end{equation}%
where $n_{a}=1$, $n_{b}=-1$ and we have introduced the notation%
\begin{equation}
B_{jn}(u)=B_{j}^{2}(n^{2}/j^{2}+u^{2})+n_{j}A_{j}B_{j}/j-A_{j}^{2}.
\label{Bjnu}
\end{equation}%
This allows us to write the expressions (\ref{pjint0}) for the interaction
forces per unit surface in another equivalent form:%
\begin{eqnarray}
p_{\mathrm{(int)}}^{(j)} &=&\frac{A_{D}\delta _{i}^{k}n_{j}}{2j}%
\sideset{}{'}{\sum}_{n=0}^{\infty }\int_{m}^{\infty }du\,u\left(
u^{2}-m^{2}\right) ^{\frac{D-3}{2}}  \notag \\
&&\times \left[ 1+(4\xi -1)\frac{n_{j}A_{j}B_{j}}{jB_{jn}(u)}\right] \frac{%
\partial }{\partial j}\ln \left\vert 1-\frac{\bar{I}_{n}^{(a)}(au)\bar{K}%
_{n}^{(b)}(bu)}{\bar{I}_{n}^{(b)}(bu)\bar{K}_{n}^{(a)}(au)}\right\vert .
\label{pjint1}
\end{eqnarray}%
This form will be used in the next section in the discussion of the relation
between the vacuum forces and the Casimir energy. Note that for Dirichlet
and Neumann scalars the second term on the square brackets vanish and the
interaction forces do not depend on the curvature coupling parameter.

Now we turn to the investigation of the vacuum interaction forces in various
limiting cases. First of all let us consider the case when the radii of
cylindrical surfaces are close to each other: $b/a-1\ll 1$. Noting that in
the limit $b\rightarrow a$ the vacuum forces diverge and in the limit under
consideration the main contribution comes from large values $n$, we can use
the uniform asymptotic expansions for the Bessel modified functions (see,
for instance, \cite{Abra64}). As the next step we introduce a new
integration variable $x=\sqrt{u^{2}-m^{2}}$ and replace the summation over $n
$ by the integration: $\sum_{n=0}^{\infty \prime }\rightarrow
\int_{0}^{\infty }dn$. Further introducing a new integration variable $n=ay$
and passing to the polar coordinates in the plane $(x,y)$ after the
integration of the angular part one finds%
\begin{equation}
p_{\mathrm{(int)}}^{(j)}\approx -\frac{2^{1-D}}{\pi ^{\frac{D}{2}}\Gamma
\left( \frac{D}{2}\right) }\int_{m}^{\infty }dt\,t^{2}(t^{2}-m^{2})^{\frac{D%
}{2}-1}\left[ \frac{(A_{a}-B_{a}t)(A_{b}-B_{b}t)}{%
(A_{a}+B_{a}t)(A_{b}+B_{b}t)}e^{2(b-a)t}-1\right] ^{-1}.  \label{pintplates}
\end{equation}%
The latter formula coincides with that for the interaction forces between
two parallel plates with Robin boundary conditions (\ref{boundcond}) on them
\cite{Rome02}. In this case the interaction forces are the same for both
plates and do not depend on the curvature coupling parameter. Note that in
the limit $a\rightarrow b$ with fixed values of the boundary coefficients
and the shell radii, the renormalized single surface parts $p_{1}^{(j)}$
remain finite while the interaction part goes to infinity. This means that
for sufficiently small distances between the boundaries the interaction term
on the right of formula (\ref{pj}) will dominate.

\begin{figure}[tbph]
\begin{center}
\begin{tabular}{cc}
\epsfig{figure=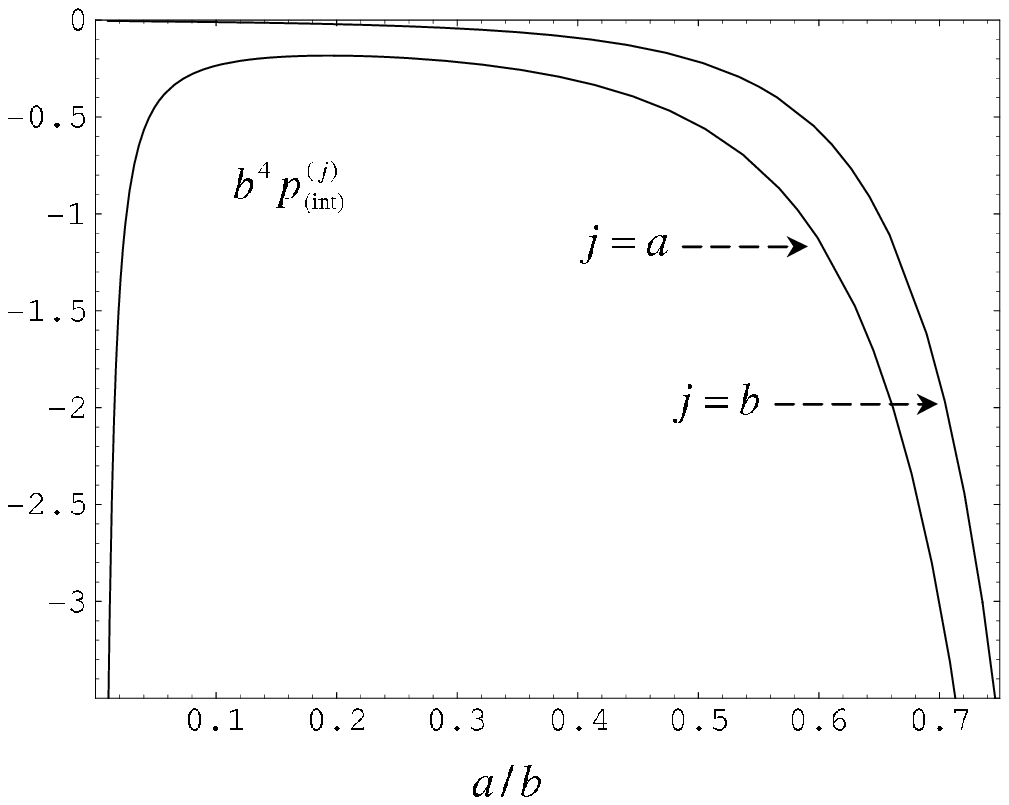,width=7.cm,height=6cm} & \quad %
\epsfig{figure=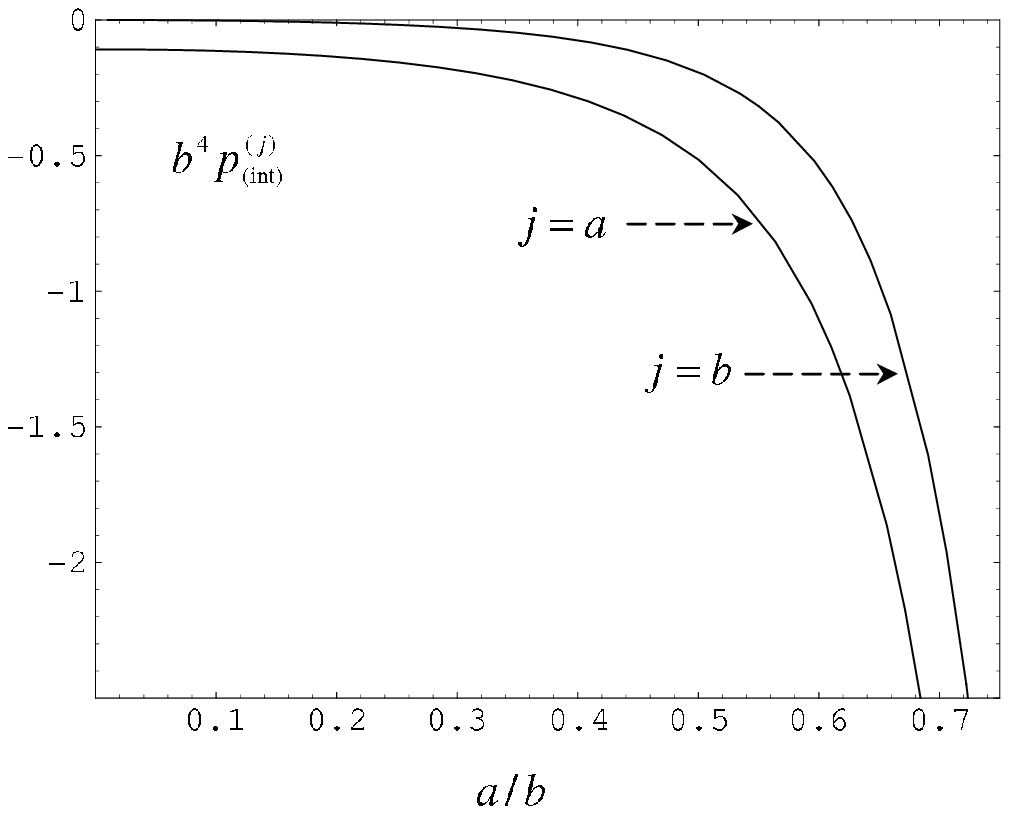,width=7.cm,height=6cm}%
\end{tabular}%
\end{center}
\caption{Vacuum interaction forces acting per unit surfaces of the
cylindrical shells as functions of $a/b$. The left panel corresponds to the
case of $D=3$ massless Dirichlet scalar and the right one is for Neumann
scalar.}
\label{fig1}
\end{figure}

For small values of the ratio $a/b$ we introduce in (\ref{pjint0}) a new
integration variable $bu=x$ and expand the integrand by using the formulae
for the Bessel modified functions for small values of the argument. For $%
A_{a},B_{a}\neq 0$ in the leading order the main contribution comes from $%
n=0 $ term and we have%
\begin{eqnarray}
p_{\mathrm{(int)}}^{(a)} &\approx &\frac{A_{D}A_{a}\xi }{aB_{a}}%
\int_{m}^{\infty }du\,u\left( u^{2}-m^{2}\right) ^{\frac{D-3}{2}}\frac{\bar{K%
}_{0}^{(b)}(bu)}{\bar{I}_{0}^{(b)}(bu)},  \label{paintfar} \\
p_{\mathrm{(int)}}^{(b)} &\approx &-\frac{A_{D}A_{a}a}{4b^{2}B_{a}}%
\int_{m}^{\infty }du\,u\left( u^{2}-m^{2}\right) ^{\frac{D-3}{2}}\frac{%
u^{2}B_{b}^{2}-4\xi A_{b}B_{b}/b-A_{b}^{2}}{\bar{I}_{0}^{(b)2}(bu)}.
\label{pbintfar}
\end{eqnarray}
In the same limit and for $B_{a}=0$ the corresponding asymptotic formulae
for the interaction forces are obtained by the replacement $\xi
A_{a}/B_{a}\rightarrow -1/[4a\ln ^{2}(a/b)]$ in formula (\ref{paintfar}) and
by the replacement $A_{a}a/B_{a}\rightarrow 1/\ln (a/b)$ in formula (\ref%
{pbintfar}). In the case of Neumann boundary condition on the surface $r=a$ (%
$A_{a}=0$) the main contribution into the vacuum interaction forces comes
from $n=0$ and $n=1$ terms with the leading behavior%
\begin{eqnarray}
p_{\mathrm{(int)}}^{(a)} &\approx &\frac{A_{D}}{2}\int_{m}^{\infty
}du\,u^{3}\left( u^{2}-m^{2}\right) ^{\frac{D-3}{2}}\sideset{}{'}{\sum}%
_{n=0}^{1}\frac{\bar{K}_{n}^{(b)}(bu)}{\bar{I}_{n}^{(b)}(bu)},
\label{paintfar1} \\
p_{\mathrm{(int)}}^{(b)} &\approx &-\frac{A_{D}a^{2}}{4b^{2}}%
\int_{m}^{\infty }du\,u^{3}\left( u^{2}-m^{2}\right) ^{\frac{D-3}{2}}%
\sideset{}{'}{\sum}_{n=0}^{1}\frac{(u^{2}+n^{2}/b^{2})B_{b}^{2}-4\xi
A_{b}B_{b}/b-A_{b}^{2}}{\bar{I}_{n}^{(b)2}(bu)}.  \label{pbintfar1}
\end{eqnarray}%
In figure \ref{fig1} we have plotted the dependence of the interaction
forces per unit surface on the ratio of the radii for the cylindrical
shells, $a/b$, for the cases of $D=3$ Dirichlet (left panel) and Neumann
(right panel) massless scalars. As we have mentioned before, the interaction
forces in these cases are attractive.
In figure \ref{fig2} we present the same graphs for the Robin boundary
conditions with the coefficients $B_{a}=0$, $B_{b}/(A_{b}b)=-0.2$. The left
panel corresponds to the minimally coupled scalar and the right one is for a
conformally coupled scalar. As we see, for the considered example the
interaction force are repulsive for small distances between the surfaces and
are attractive for large distances. This provides a possibility for the
stabilization of the radii by the vacuum forces. However, it should be noted
that to make reliable predictions regarding quantum stabilization, the
renormalized single shell parts $p_{1}^{(j)}$ also should be taken into
account.

\begin{figure}[tbph]
\begin{center}
\begin{tabular}{cc}
\epsfig{figure=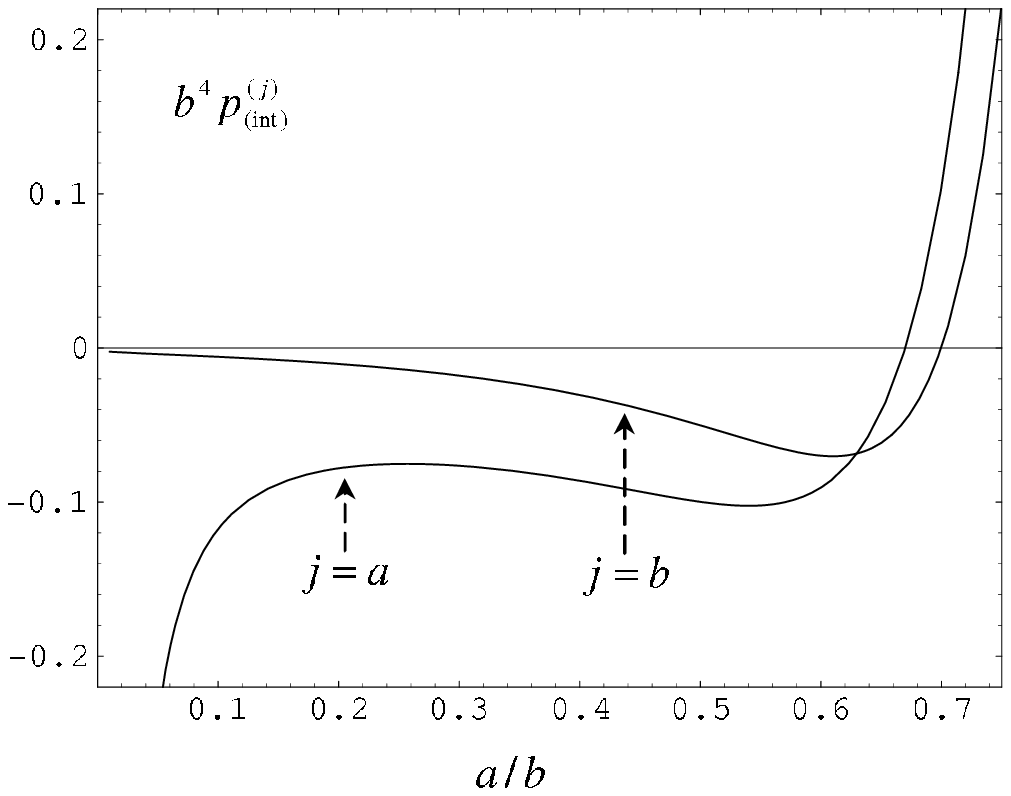,width=7.cm,height=6cm} & \quad %
\epsfig{figure=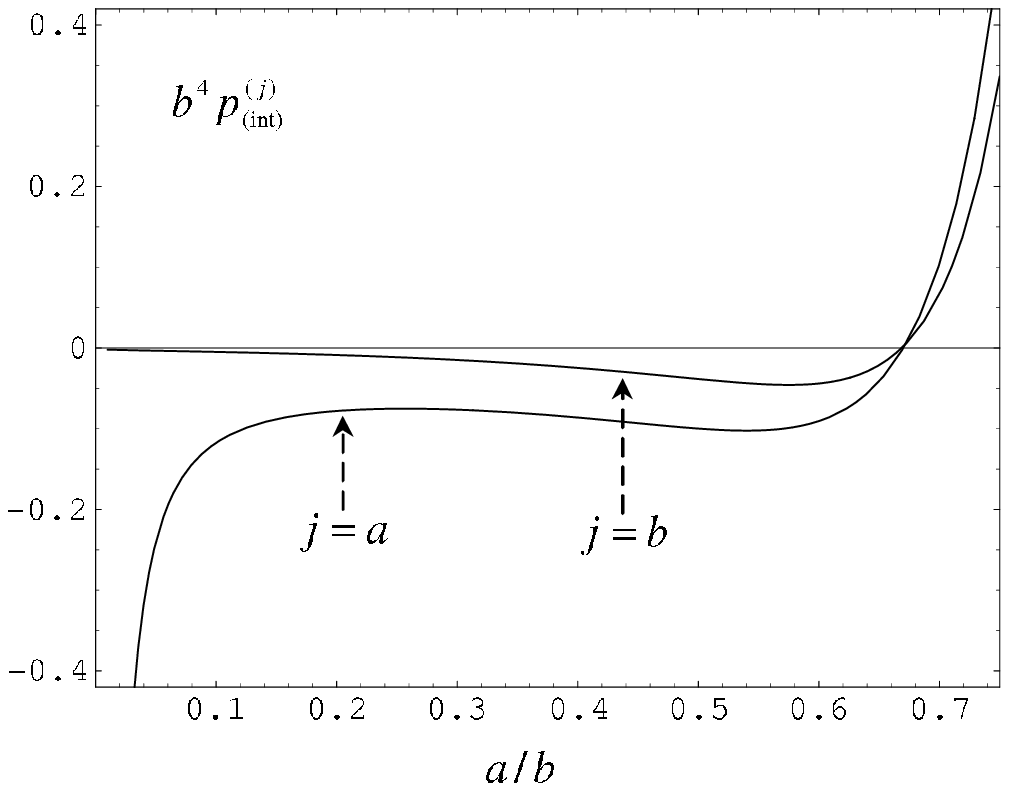,width=7.cm,height=6cm}%
\end{tabular}%
\end{center}
\caption{The same as in figure \protect\ref{fig1} for minimally (left panel)
and conformally (right panel) coupled Robin scalars with $B_{a}=0$, $%
B_{b}/(A_{b}b)=-0.2$. }
\label{fig2}
\end{figure}

\section{Casimir energy}

\label{sec:toten}

\begin{figure}[tbph]
\begin{center}
\epsfig{figure=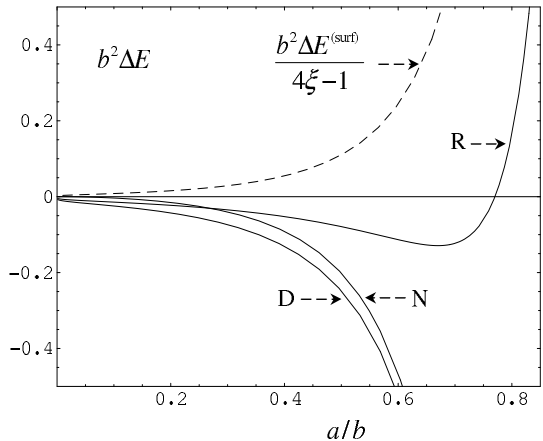,width=7.5cm,height=6.cm}
\end{center}
\caption{Interaction parts of the total and surface energies as functions of
$a/b$. The graphs are plotted for $D=3$ massless scalar fields with
Dirichlet (D) and Neumann (D) boundary conditions and in the case of Robin
(R) boundary condition with $B_{a}=0$, $B_{b}/(A_{b}b)=-0.2$.}
\label{fig3}
\end{figure}

In this section we consider the total vacuum energy for the configuration of
two coaxial cylindrical boundaries. In the region between the boundaries the
total vacuum energy per unit hypersurface in the axial direction is the sum
of the zero-point energies of elementary oscillators:
\begin{equation}
E_{a\leqslant r\leqslant b}=\int \frac{d^{N}{\mathbf{k}}}{(2\pi )^{N}}%
\sideset{}{'}{\sum}_{n=0}^{\infty }\sum_{l=1}^{\infty }(k^{2}+m^{2}+\sigma
_{n,l}^{2}/a^{2})^{1/2}.  \label{toten1}
\end{equation}%
The expression on the right is divergent and to deal with this divergence we
take the zeta function approach (for the application of the zeta function
technique to the calculations of the Casimir energy see \cite{Kirs02} and
references therein). We consider the related zeta function%
\begin{equation}
\zeta (s)=\mu ^{s+1}\int \frac{d^{N}{\mathbf{k}}}{(2\pi )^{N}}%
\sideset{}{'}{\sum}_{n=0}^{\infty }\sum_{l=1}^{\infty }(k^{2}+\sigma
_{n,l}^{2}/a^{2}+m^{2})^{-s/2},  \label{zeta}
\end{equation}%
where the parameter $\mu $ with dimension of mass is introduced by
dimensional reasons. Evaluating the integral over ${\mathbf{k}}$ we present
this function in the form%
\begin{equation}
\zeta (s)=\frac{\mu ^{s+1}}{(4\pi )^{\frac{N}{2}}}\frac{\Gamma \left( \frac{%
s-N}{2}\right) }{\Gamma \left( \frac{s}{2}\right) a^{N-s}}\sideset{}{'}{\sum}%
_{n=0}^{\infty }\zeta _{n}\left( s-N\right) ,  \label{zeta1}
\end{equation}%
with the partial zeta function%
\begin{equation}
\zeta _{n}\left( s\right) =\sum_{l=1}^{\infty }(\sigma
_{n,l}^{2}+m^{2}a^{2})^{-s/2}.  \label{zetan}
\end{equation}%
We need to perform the analytic continuation of the sum on the right of (\ref%
{zeta1}) to the neighborhood of $s=-1$. An immediate consequence of the
Cauchy's formula for the residues of a complex function is the expression%
\begin{equation}
\zeta _{n}\left( s\right) =\frac{1}{2\pi i}\int_{C}dz%
\,(z^{2}+m^{2}a^{2})^{-s/2}\frac{\partial }{\partial z}\ln C_{n}^{ab}(\eta
,z),  \label{zetanint1}
\end{equation}%
where $C$ is a closed counterclockwise contour in the complex $z$ plane
enclosing all zeros $\sigma _{n,l}$. We assume that this contour is made of
a large semicircle (with radius tending to infinity) centered at the origin
and placed to its right, plus a straight part overlapping the imaginary axis
and avoiding the points $\pm iam$ by small semicircles in the left
half-plane. When the radius of the large semicircle tends to infinity the
corresponding contribution into $\zeta _{n}\left( s\right) $ vanishes for ${%
\mathrm{Re}}\,s>1$. Let us denote by $C^{1}$ and $C^{2}$ the upper and lower
halves of the contour $C$. The integral on the right of Eq. (\ref{zetanint1}%
) can be presented in the form%
\begin{eqnarray}
\zeta _{n}\left( s\right) &=&\frac{1}{2\pi i}\int_{C}dz%
\,(z^{2}+m^{2}a^{2})^{-s/2}\frac{\partial }{\partial z}\ln \left[ z^{-n}\bar{%
J}_{n}^{(b)}(\eta z)\right]  \notag \\
&&+\frac{1}{2\pi i}\sum_{\alpha =1,2}\int_{C^{\alpha
}}dz\,(z^{2}+m^{2}a^{2})^{-s/2}\frac{\partial }{\partial z}\ln \left[ z^{n}%
\bar{H}_{n}^{(\alpha a)}(z)\right]  \notag \\
&&+\frac{1}{2\pi i}\sum_{\alpha =1,2}\int_{C^{\alpha
}}dz\,(z^{2}+m^{2}a^{2})^{-s/2}\frac{\partial }{\partial z}\ln \left[ 1-%
\frac{\bar{J}_{n}^{(a)}(z)\bar{H}_{n}^{(\alpha b)}(\eta z)}{\bar{H}%
_{n}^{(\alpha a)}(z)\bar{J}_{n}^{(b)}(\eta z)}\right] ,  \label{zetanint2}
\end{eqnarray}%
where $H_{n}^{(\alpha )}(z)$ are the Hankel functions. After parameterizing
the integrals over imaginary axis we see that the parts of the integrals
over $(0,\pm ima)$ cancel and we arrive at the expression%
\begin{eqnarray}
\zeta _{n}\left( s\right) &=&\frac{1}{\pi }\sin \frac{\pi s}{2}%
\int_{ma}^{\infty }dz\,(z^{2}-m^{2}a^{2})^{-s/2}\frac{\partial }{\partial z}%
\Bigg\{\ln \left[ z^{-n}\bar{I}_{n}^{(b)}(\eta z)\right]  \notag \\
&&+\ln \left[ z^{n}\bar{K}_{n}^{(a)}(z)\right] +\ln \left( 1-\frac{\bar{I}%
_{n}^{(a)}(z)\bar{K}_{n}^{(b)}(\eta z)}{\bar{K}_{n}^{(a)}(z)\bar{I}%
_{n}^{(b)}(\eta z)}\right) \Bigg\} .  \label{zetanint3}
\end{eqnarray}%
The integral with the last term in square brackets on the right of this
formula is finite at $s=-(N+1)$ and vanishes in the limits $a\rightarrow 0$
or $b\rightarrow \infty $. The integrals with the first and second terms in
the square brackets correspond to the partial zeta functions for the region
inside a cylindrical shell with radius $b$ and for the region outside a
cylindrical shell with radius $a$, respectively. As a result the total
energy in the region $a\leqslant r\leqslant b$ is presented in the form%
\begin{equation}
E_{a\leqslant r\leqslant b}=\zeta (s)|_{s=-1}=E_{r\geqslant
a}^{(a)}+E_{r\leqslant b}^{(b)}+\Delta E,  \label{Eab}
\end{equation}%
where $E_{r\geqslant a}^{(a)}$ ($E_{r\leqslant b}^{(b)}$) is the vacuum
energy for the region outside (inside) a cylindrical shell with radius $a$ ($%
b$) and the interference term is given by the formula%
\begin{eqnarray}
\Delta E &=&-\frac{(4\pi )^{\frac{1-D}{2}}}{\Gamma (\frac{D+1}{2})}%
\sideset{}{'}{\sum}_{n=0}^{\infty }\int_{m}^{\infty }du\,(u^{2}-m^{2})^{%
\frac{D-1}{2}}\frac{\partial }{\partial u}\ln \left\vert 1-\frac{\bar{I}%
_{n}^{(a)}(au)\bar{K}_{n}^{(b)}(bu)}{\bar{K}_{n}^{(a)}(au)\bar{I}%
_{n}^{(b)}(bu)}\right\vert  \notag \\
&=&\pi A_{D}\sideset{}{'}{\sum}_{n=0}^{\infty }\int_{m}^{\infty
}du\,u(u^{2}-m^{2})^{\frac{D-3}{2}}\ln \left\vert 1-\frac{\bar{I}%
_{n}^{(a)}(au)\bar{K}_{n}^{(b)}(bu)}{\bar{K}_{n}^{(a)}(au)\bar{I}%
_{n}^{(b)}(bu)}\right\vert  \label{DeltaEn}
\end{eqnarray}%
To obtain the first of these formulae from the corresponding zeta function
we have used the relation $\Gamma (x)\sin \pi x=\pi /\Gamma (1-x)$ for the
gamma function. The interaction part of the vacuum energy (\ref{DeltaEn}) is
negative for Dirichlet or Neumann boundary conditions and positive for
Dirichlet boundary condition on one shell and Neumann boundary condition on
the another. By the way similar to that used before for the case of the
interaction forces it can be seen that in the limit $a,b\rightarrow \infty $
for fixed $b-a$ the corresponding result is obtained for parallel plates. In
the limit $a\rightarrow 0$ for $A_{a},B_{a}\neq 0$ the main contribution
into the interaction part of the vacuum energy comes from $n=0$ term. By
using the expansions for the Bessel modified functions for small values of
the argument to the leading order one finds%
\begin{equation}
\Delta E\approx A_{D}\frac{\pi A_{a}a}{2B_{a}}\int_{m}^{\infty
}du\,u(u^{2}-m^{2})^{\frac{D-3}{2}}\frac{\bar{K}_{0}^{(b)}(bu)}{\bar{I}%
_{0}^{(b)}(bu)}.  \label{DeltaEsmalla}
\end{equation}%
In the same limit and for Dirichlet boundary condition on the inner
cylinder, $B_{a}=0$, the leading behavior for $\Delta E$ is obtained from (%
\ref{DeltaEsmalla}) by the replacement $A_{a}a/B_{a}\rightarrow \ln (a/b)$.
In the case of Neumann boundary condition on $r=a$, the main contribution
comes from $n=0$ and $n=1$ terms and $\Delta E$ vanishes as $a^{2}$:%
\begin{equation*}
\Delta E\approx A_{D}\frac{\pi a^{2}}{2}\int_{m}^{\infty
}du\,u^{3}(u^{2}-m^{2})^{\frac{D-3}{2}}\sideset{}{'}{\sum}_{n=0}^{1}\frac{%
\bar{K}_{n}^{(b)}(bu)}{\bar{I}_{n}^{(b)}(bu)}
\end{equation*}
In the limit $b\rightarrow \infty $ and for a massless scalar field we have
an asymptotic behavior with the leading term coming from $n=0$ summand:%
\begin{equation}
\Delta E\approx \frac{\pi A_{D}}{2b^{D-1}\ln (a/b)}\int_{0}^{\infty
}du\,u^{D-2}\frac{K_{0}(u)}{I_{0}(u)},  \label{DeltaElargebm0}
\end{equation}%
assuming that $A_{a},A_{b}\neq 0$. For $A_{a}=0$ and $A_{b}\neq 0$ the main
contribution into the interaction part of the vacuum energy comes from $n=0$
and $n=1$ terms:%
\begin{equation}
\Delta E\approx \frac{\pi a^{2}A_{D}}{2b^{D+1}}\int_{0}^{\infty }du\,u^{D}%
\sideset{}{'}{\sum}_{n=0}^{1}\frac{K_{n}(bu)}{I_{n}(bu)}.
\label{DeltaElargebm01}
\end{equation}%
For the Neumann boundary condition on the outer cylinder, $A_{b}=0$, in the
integrands of (\ref{DeltaElargebm0}) and (\ref{DeltaElargebm01}) instead of
ratio of the Bessel modified functions the ratio of their derivatives
stands. For a massive field and large values for the radius of the outer
cylinder, under the condition $mb\gg 1$ the main contribution into the
integral over $u$ in Eq. (\ref{DeltaEn}) comes from the lower limit of the
integral. By using the asymptotic formulae for the Bessel modified function
for large values of the argument, to the leading order we find
\begin{equation}
\Delta E\approx -\frac{2^{1-D}m^{\frac{D-1}{2}}}{\pi ^{\frac{D-3}{2}}b^{%
\frac{D-1}{2}}}\frac{A_{b}-B_{b}m}{A_{b}+B_{b}m}e^{-2mb}\sideset{}{'}{\sum}%
_{n=0}^{\infty }\frac{\bar{I}_{n}^{(a)}(am)}{\bar{K}_{n}^{(a)}(am)},
\label{DeltaElargebmnon}
\end{equation}%
and the interaction part of the vacuum energy is exponentially suppressed.
In figure \ref{fig3} we have plotted the dependence of the interaction parts
(full curves) in the total vacuum energy on the ratio $a/b$ for $D=3$
massless scalar fields with Dirichlet, Neumann and Robin boundary
conditions. For the Robin case we have chosen the parameters in the boundary
conditions as $B_{a}=0$, $B_{b}/(A_{b}b)=-0.2$.

To obtain the total Casimir energy, $E$, we need to add to the energy in the
region between the shells, given by Eq. (\ref{DeltaEn}), the energies coming
from the regions $r\leqslant a$ and $r\geqslant b$. As a result one receives%
\begin{equation}
E=\sum_{j=a,b}E^{(j)}+\Delta E,  \label{Etot}
\end{equation}%
where $E^{(j)}$ is the Casimir energy for a single cylindrical shell with
the radius $j$. The latter is investigated in Ref. \cite{Rome01} as a
function on the ratio of coefficients in Robin boundary condition. In
particular, for Dirichlet and Neumann massless scalars one has the Casimir
energies $E_{D}^{(j)}=6.15\cdot 10^{-4}j^{-2}$ and $E_{N}^{(j)}=-1.42\cdot
10^{-2}j^{-2}$ (see \cite{Gosd98}). Note that in the total vacuum energy
single boundary parts dominate for small values of the ratio $a/b$ and the
interaction part is dominant for $a/b\lesssim 1$. In particular, combining
the results for the single surface energies with the graphs from figure \ref%
{fig3}, we see that the total Casimir energy for a massless Dirichlet scalar
in the geometry of two cylindrical shells is positive for small values $a/b$
and is negative for $a/b\lesssim 1$. For Neuamnn case the vacuum energy is
negative for all values $a/b$ and has a maximum for some intermediate value
of this ratio.

The total volume energy in the region $a\leqslant r\leqslant b$ is derived
by the integration of ${_{0}^{0}}$-component of the volume energy-momentum
tensor over this region:
\begin{equation}
E^{{\mathrm{(vol)}}}=2\pi \int_{a}^{b}dr\,r\langle 0|T_{0}^{0}|0\rangle .
\label{Evol}
\end{equation}%
Substituting mode-sum expansion (\ref{W2}) into formula (\ref{emtvev1}),
after the integration for the volume part of the vacuum energy we obtain%
\begin{equation}
E^{{\mathrm{(vol)}}}=E_{a\leqslant r\leqslant b}-\pi (4\xi -1)\sum_{j=a,b}%
\frac{jA_{j}}{B_{j}}\langle 0|\varphi ^{2}|0\rangle _{r=j}.  \label{Evol1}
\end{equation}%
As we see this energy differs from the total Casimir energy (\ref{toten1})
(see Ref. \cite{Saha04} for the discussion in general case of bulk and
boundary geometries). This difference is due to the presence of the surface
energy located on the bounding surfaces. By using the standard variational
procedure, in Ref. \cite{Saha04} it has been shown that the energy-momentum
tensor for a scalar field on manifolds with boundaries in addition to the
bulk part contains a contribution located on the boundary. For an arbitrary
smooth boundary $\partial M_{s}$ with the inward-pointing unit normal vector
$n^{l}$, the surface part of the energy-momentum tensor is given by the
formula
\begin{equation}
T_{ik}^{\mathrm{(surf)}}=\delta (x;\partial M_{s})\tau _{ik}
\label{Ttausurf}
\end{equation}%
with
\begin{equation}
\tau _{ik}=\xi \varphi ^{2}K_{ik}-(2\xi -1/2)h_{ik}\varphi n^{l}\nabla
_{l}\varphi ,  \label{tausurf}
\end{equation}%
and the 'one-sided' delta-function $\delta (x;\partial M_{s})$ locates this
tensor on $\partial M_{s}$. In Eq. (\ref{tausurf}), $K_{ik}$ is the
extrinsic curvature tensor of the boundary $\partial M_{s}$ and $h_{ik}$ is
the corresponding induced metric. In the region between the cylindrical
surfaces for separate boundaries one has $K_{ik}^{(j)}=-jn_{j}\delta
_{i}^{2}\delta _{k}^{2}$, $j=a,b$. Now substituting the eigenfunctions into
the corresponding mode-sum formula and using the boundary conditions, for
the surface energy-momentum tensor on the boundary $r=j$ one finds
\begin{equation}
\langle 0|\tau _{i}^{(j)k}|0\rangle =\left[ -\xi jn_{j}\delta _{i}^{2}\delta
_{2}^{k}+(2\xi -1/2)\delta _{i}^{k}A_{j}/B_{j}\right] \langle 0|\varphi
^{2}|0\rangle _{r=j},  \label{tauj}
\end{equation}%
for $i,k=0,1,\ldots ,D$, and $\langle 0|\tau _{1}^{(j)1}|0\rangle =0$. From (%
\ref{tauj}) for the surface energy we obtain the formula
\begin{equation}
E^{{\mathrm{(surf)}}}=2\pi \int_{a}^{b}dr\,r\langle 0|T_{0}^{\mathrm{(surf)}%
0}|0\rangle =\pi (4\xi -1)\sum_{j=a,b}\frac{jA_{j}}{B_{j}}\langle 0|\varphi
^{2}|0\rangle _{r=j},  \label{Esurf}
\end{equation}%
which, in accordance with (\ref{Evol1}), exactly coincides with the
difference between the total and volume energies. Of course, the VEVs\ of
the field square on the right of this formula and, hence, the surface
energy-momentum tensor diverge and need some regularization with further
renormalization. Note that due to the surface divergences the subtraction of
the boundary-free part is not sufficient to obtain the finite result and
additional renormalization procedure is needed. In particular, the
generalized zeta function method is in general very powerful to give
physical meaning to the divergent quantities. However, in this paper we will
not go into the details of the renormalization for the surface
energy-momentum tensor for a single cylindrical boundary. This investigation
will be presented in the forthcoming paper \cite{SahTarl}. Here we note that
after the subtraction of the boundary-free part the remained divergences in
the surface energy-momentum tensor are the same as those for a single
surfaces when the second surface is absent. The additional parts induced by
the presence of the second surface are finite and can be obtained by using
the representation of the VEV for the field square given by formula (\ref%
{phi21}). By taking into account the relation (\ref{Gnuu}), the surface
energy on the boundary $r=j$ is presented in the form%
\begin{equation}
E_{j}^{{\mathrm{(surf)}}}=E_{1j}^{{\mathrm{(surf)}}}+\Delta E_{j}^{{\mathrm{%
(surf)}}},  \label{Esurf1}
\end{equation}%
where $E_{1j}^{{\mathrm{(surf)}}}$ is the surface energy for a single
cylindrical boundary with radius $j$ when the second boundary is absent and
the term%
\begin{equation}
\Delta E_{j}^{{\mathrm{(surf)}}}=-\pi (4\xi -1)A_{D}\frac{A_{j}B_{j}}{j}%
\sideset{}{'}{\sum}_{n=0}^{\infty }\int_{m}^{\infty }du\,u\left(
u^{2}-m^{2}\right) ^{\frac{D-3}{2}}\Omega _{jn}(au,bu)  \label{DeltaEjsurf}
\end{equation}%
is induced by the presence of the second boundary. The latter is finite for
all non-zero intersurface distances. Note that on the base of relation (\ref%
{ident2}) it can also be written in the form%
\begin{eqnarray}
\Delta E_{j}^{{\mathrm{(surf)}}} &=&-\pi A_{D}(4\xi -1)\sideset{}{'}{\sum}%
_{n=0}^{\infty }\int_{m}^{\infty }du\,u\left( u^{2}-m^{2}\right) ^{\frac{D-3%
}{2}}  \notag \\
&&\times \frac{n_{j}A_{j}B_{j}}{B_{jn}(u)}\frac{\partial }{\partial j}\ln
\left\vert 1-\frac{\bar{I}_{n}^{(a)}(au)\bar{K}_{n}^{(b)}(bu)}{\bar{I}%
_{n}^{(b)}(bu)\bar{K}_{n}^{(a)}(au)}\right\vert ,  \label{DeltaEjsurf1}
\end{eqnarray}%
where $B_{jn}(u)$ \ is defined by Eq. (\ref{Bjnu}). Now, by using formulae (%
\ref{DeltaEvol}), (\ref{DeltaEn}), (\ref{DeltaEjsurf1}) it can be explicitly
checked the relation%
\begin{equation}
\Delta E=\Delta E^{\mathrm{(vol)}}+\sum_{j=a,b}\Delta E_{j}^{{\mathrm{(surf)}%
}}  \label{Deltarel}
\end{equation}%
for the interaction parts of the separate energies. In figure \ref{fig3} we
have plotted the interaction part of the surface energy (dashed curve) for
the case of $D=3$ massless Robin scalar with $B_{a}=0$, $B_{b}/(A_{b}b)=-0.2$
as a function on the ratio $a/b$. For the considered example this energy is
located on the surface $r=b-0$ of the outer cylinder. The surface energies
for Dirichlet and Neumann scalars vanish.

Now let us explicitly check that for the interaction parts the standard
energy balance equation is satisfied. We expect that in the presence of the
surface energy this equation will be in the form
\begin{equation}
dE=-pdV+\sum_{j=a,b}\frac{E_{j}^{{\mathrm{(surf)}}}}{2\pi j}dS^{(j)},
\label{enbalance}
\end{equation}%
where $V=\pi (b^{2}-a^{2})$ and $S^{(j)}=2\pi j$ are the volume and surface
area per unit hypersurface in the axial direction. In Eq. (\ref{enbalance}),
$p$ is the perpendicular vacuum stress on the boundary and is determined by
the vacuum expectation value of the ${}_{1}^{1}$-component of the bulk
energy-momentum tensor: $p=-\langle 0|T_{1}^{1}|0\rangle $. From equation (%
\ref{enbalance}) one obtains
\begin{equation}
\frac{\partial E}{\partial j}=2\pi jn_{j}p^{(j)}+E_{j}^{{\mathrm{(surf)}}}/j,
\label{dEdzj}
\end{equation}%
with $p^{(j)}$ being the perpendicular vacuum stress on the boundary $r=j$.
Assuming that relation (\ref{dEdzj}) is satisfied for the single boundary
parts, for the interference parts we find
\begin{equation}
\frac{\partial \Delta E}{\partial j}=2\pi jn_{j}p_{{\mathrm{(int)}}%
}^{(j)}+\Delta E_{j}^{{\mathrm{(surf)}}}/j.  \label{dDeltaEdzj}
\end{equation}%
Now by taking into account expressions (\ref{pjint1}), (\ref{DeltaEn}), (\ref%
{DeltaEjsurf1}) for the separate terms in this formula and integrating by
part in (\ref{DeltaEn}), we see that this relation indeed takes place.
Hence, we have explicitly checked that the vacuum energies and effective
pressures on the boundaries obey the standard energy balance equation. Note
that here the role of the surface energy is crucial and the vacuum forces
acting on the boundary evaluated from the bulk stress tensor [determined by $%
p_{{\mathrm{(int)}}}^{(j)}$], in general, can not be obtained by a simple
differentiation of the total vacuum energy. The second term on the right of
formula (\ref{dDeltaEdzj}) corresponds to the additional pressure acting on
the curved boundary. Noting that $\Delta E_{j}^{{\mathrm{(surf)}}}/(2\pi j)$
is the corresponding surface energy density, we see that this pressure is
determined by Laplace formula. The total pressure on the boundary evaluated
as the sum of bulk and surface parts is related to the total vacuum energy
by standard formula and does not depend on the curvature coupling parameter.

\section{Conclusion}

\label{sec:Conc}

In the present paper, we have investigated the one-loop quantum vacuum
effects produced by two coaxial cylindrical shells in the $(D+1)$%
-dimensional Minkowski spacetime. The case of a massive scalar field with
general curvature coupling parameter and satisfying the Robin boundary
conditions on the boundaries is considered. To derive formulae for the VEVs
of the field operator squared and the energy-momentum tensor, we first
construct the positive frequency Wightman function. This function is also
important in considerations of the response of a particle detector at a
given state of motion through the vacuum under consideration \cite{Birr82}.
The application of a variant of the generalized Abel-Plana formula to the
mode-sum over zeros of the combinations of the cylindrical functions allowed
us to extract the parts due to a single cylindrical boundary and to present
the second boundary induced parts in terms of the exponentially convergent
integrals. For the exterior and interior regions of a single cylindrical
shell the Wightman functions are given by formulae (\ref{Wa}) and (\ref{Wb})
respectively. The second boundary induced parts are presented by the last
terms on the right of formulae (\ref{W4}) and (\ref{W5}). The VEVs of the
field square and the energy-momentum tensor are obtained by the evaluation
of the Wightman function and the combinations of its derivatives in the
coincidence limit of arguments. In both cases the expectation values are
presented as the sum of single boundary induced and interference terms. The
surface divergences in the VEVs of the local observables are contained in
the single boundary parts and the interference parts are finite on both
boundaries. In particular, the integrals in the corresponding formulae are
exponentially convergent and they are useful for numerical evaluations. Due
to the presence of boundaries the vacuum stresses in radial, azimuthal and
axial directions are anisotropic. For the axial stress and the energy
density we have standard relation for the unbounded vacuum. We have
considered various limiting cases of the formulae for the interference
parts. In particular, in the limit $a,b\rightarrow \infty $ for a fixed
value $b-a$, we recover the result for the geometry of two parallel Robin
plates on the Minkowski background. The vacuum forces acting on boundaries
are considered in section \ref{sec:forces}. These forces contain two terms.
The first ones are the forces acting on a single surface then the second
boundary is absent. Due to the well-known surface divergences in the VEVs of
the energy-momentum tensor these forces are infinite and need an additional
regularization with further renormalization. The another terms in the vacuum
forces are finite and are induced by the presence of the second boundary.
They correspond to the interaction forces between the boundaries and are
determined by formula (\ref{pjint0}) or equivalently by formula (\ref{pjint1}%
). For the Dirichlet and Neumann scalars these forces are always attractive
and they are repulsive for the mixed Dirichlet-Neumann case. In the case of
general Robin scalar the interaction forces can be both attractive or
repulsive in dependence of the coefficients in the boundary conditions and
the distance between the boundaries. As an illustration, in figure \ref{fig2}
we present an example when the interaction forces are repulsive for small
distances and are attractive for large distances. This provides a
possibility for the stabilization of the radii by vacuum forces. However, it
should be noted that for the reliable predictions regarding quantum
stabilization, the renormalized single shell parts also should be taken into
account. In section \ref{sec:toten} we consider the total vacuum energy in
the region between the cylindrical surfaces, evaluated as the sum of the
zero-point energies for elementary oscillators. It is argued that this
energy differs from the energy, obtained by the integration of the volume
energy density over the region between the boundaries. We show that this
difference is due to the presence of the surface energy located on the
bounding surfaces. Further for the evaluation of the total and surface
energies we use the zeta function technique. They are presented as the sum
of single boundary and interaction parts. The latter are given by formula (%
\ref{DeltaEn}) for the total vacuum energy and by formula (\ref{DeltaEjsurf1}%
) for the surface energy and are finite for all nonzero values of the
intersurface separation. In the total vacuum energy single boundary parts
dominate for small values of the ratio $a/b$ and the interaction part is
dominant for $a/b\lesssim 1$. For an arbitrary number of spatial dimensions
and independent on the value of the mass, the interaction part of the vacuum
energy is negative for Dirichlet or Neumann boundary conditions and is
positive for Dirichlet boundary condition on one shell and Neumann boundary
condition on the another. Further, we have shown that the induced vacuum
densities and vacuum effective pressures on the cylindrical surfaces satisfy
the energy balance equation (\ref{enbalance}) with the inclusion of the
surface terms, which can also be written in the form (\ref{dDeltaEdzj}).

\section*{Acknowledgements}

AAS was supported by the Armenian Ministry of Education and Science Grant
No. 0124 and in part by PVE/CAPES program.

\end{document}